\documentclass[aps,prl,twocolumn,amsmath,amssymb,nofootinbib,superscriptaddress]{revtex4-2}
\usepackage{times}
\usepackage{braket}
\usepackage[pdftex]{graphicx}
\usepackage{dcolumn}
\usepackage{bm}
\usepackage{bbm}
\usepackage{amsmath}
\usepackage{indentfirst}
\usepackage[colorlinks]{hyperref}
\usepackage[dvipsnames]{xcolor}
\usepackage{xcolor}

\colorlet{shadecolor}{gray!40}

\usepackage{subfigure}
\usepackage{algorithm}
\usepackage{algorithmicx}
\usepackage{algpseudocode}
\usepackage{amsmath}
\usepackage{verbatim}
\usepackage{makecell}
\usepackage[normalem]{ulem}
\usepackage{setspace}
\usepackage{float}
\usepackage{booktabs}
\usepackage{threeparttable}
\usepackage{lineno}
\begin{document}


\title{Demonstrating a universal logical gate set {in error-detecting surface codes} on a superconducting quantum processor}

\author{Jiaxuan Zhang}
\affiliation{Key Laboratory of Quantum Information, Chinese Academy of Sciences, School of Physics, University of Science and Technology of China, Hefei, Anhui, 230026, P. R. China}
\affiliation{CAS Center For Excellence in Quantum Information and Quantum Physics, University of Science and Technology of China, Hefei, Anhui, 230026, P. R. China}

\author{Zhao-Yun Chen}
\affiliation{Institute of Artificial Intelligence, Hefei Comprehensive National Science Center, Hefei, Anhui, 230088, P. R. 
 China}

\author{Yun-Jie Wang}
\affiliation{Institute of the Advanced Technology, University of Science and Technology of China, Hefei, Anhui, 230088, P. R. China}

\author{Bin-Han Lu}
\affiliation{Key Laboratory of Quantum Information, Chinese Academy of Sciences, School of Physics, University of Science and Technology of China, Hefei, Anhui, 230026, P. R. China}
\affiliation{CAS Center For Excellence in Quantum Information and Quantum Physics, University of Science and Technology of China, Hefei, Anhui, 230026, P. R. China}

\author{Hai-Feng Zhang}
\affiliation{Key Laboratory of Quantum Information, Chinese Academy of Sciences, School of Physics, University of Science and Technology of China, Hefei, Anhui, 230026, P. R. China}
\affiliation{CAS Center For Excellence in Quantum Information and Quantum Physics, University of Science and Technology of China, Hefei, Anhui, 230026, P. R. China}

\author{Jia-Ning Li}
\affiliation{Key Laboratory of Quantum Information, Chinese Academy of Sciences, School of Physics, University of Science and Technology of China, Hefei, Anhui, 230026, P. R. China}
\affiliation{CAS Center For Excellence in Quantum Information and Quantum Physics, University of Science and Technology of China, Hefei, Anhui, 230026, P. R. China}

\author{Peng Duan}
\email{pengduan@ustc.edu.cn}
\affiliation{Key Laboratory of Quantum Information, Chinese Academy of Sciences, School of Physics, University of Science and Technology of China, Hefei, Anhui, 230026, P. R. China}
\affiliation{CAS Center For Excellence in Quantum Information and Quantum Physics, University of Science and Technology of China, Hefei, Anhui, 230026, P. R. China}

\author{Yu-Chun Wu}
\email{wuyuchun@ustc.edu.cn}
\affiliation{Key Laboratory of Quantum Information, Chinese Academy of Sciences, School of Physics, University of Science and Technology of China, Hefei, Anhui, 230026, P. R. China}
\affiliation{CAS Center For Excellence in Quantum Information and Quantum Physics, University of Science and Technology of China, Hefei, Anhui, 230026, P. R. China}
\affiliation{Institute of Artificial Intelligence, Hefei Comprehensive National Science Center, Hefei, Anhui, 230088, P. R. China}

\author{Guo-Ping Guo}
\email{gpguo@ustc.edu.cn}
\affiliation{Key Laboratory of Quantum Information, Chinese Academy of Sciences, School of Physics, University of Science and Technology of China, Hefei, Anhui, 230026, P. R. China}
\affiliation{CAS Center For Excellence in Quantum Information and Quantum Physics, University of Science and Technology of China, Hefei, Anhui, 230026, P. R. China}
\affiliation{Institute of Artificial Intelligence, Hefei Comprehensive National Science Center, Hefei, Anhui, 230088, P. R. China}
\affiliation{Origin Quantum Computing Hefei, Anhui 230026, P. R. China}
\date{\today}


\begin{abstract}
Fault-tolerant quantum computing (FTQC) is essential for achieving large-scale practical quantum computation. Implementing arbitrary FTQC requires the execution of a universal gate set on logical qubits, which is highly challenging. Particularly, in the superconducting system, two-qubit gates on surface code logical qubits have not been realized. Here, we experimentally implement a logical CNOT gate along with arbitrary single-qubit rotation gates on distance-2 surface codes using the superconducting quantum processor \textit{Wukong}, thereby demonstrating a universal logical gate set. {In the experiment, we demonstrate the transversal CNOT gate on a two-dimensional topological processor based on a tailored encoding circuit, at the cost of removing the ancilla qubits required for stabilizer measurements. Furthermore, we fault-tolerantly prepare logical Bell states and observe a violation of CHSH inequality, confirming the entanglement between logical qubits.} Using the logical CNOT gate and an ancilla logical state, arbitrary single-qubit rotation gates are realized through gate teleportation. All logical gates are characterized on a complete state set and their fidelities are evaluated by logical Pauli transfer matrices. {The demonstration of a universal logical gate set and the entangled logical states highlights significant aspects of FTQC on superconducting quantum processors.}
\end{abstract}

\maketitle

\section{Introduction}
Quantum computing holds the promise to accelerate classical computing in various applications such as large number factorization~\cite{365700}, quantum simulation~\cite{freedman2002simulation}, and machine learning~\cite{biamonte2017quantum}. However, physical qubits are typically very fragile and are easily disturbed by environmental noise. To address the noise issues in large-scale quantum computing, quantum error correction techniques have been proposed, which introduce redundant information and encode quantum states onto logical qubits to ensure fault tolerance~\cite{preskill1998reliable,gottesman1997stabilizer,nielsen2010quantum}.

In recent years, multiple experiments across various quantum computing platforms have demonstrated the memory of quantum information on logical qubits. These experiments are based on hardware systems encompassing superconducting~\cite{andersen2020repeated,google2021exponential,krinner2022realizing,PhysRevLett.129.030501,google2023suppressing,hetenyi2024creating}, ion trap~\cite{da2024demonstration,physrevx.11.041058}, neutral atom~\cite{bluvstein2024logical}, and other systems~\cite{campagne2020quantum,gertler2021protecting,sivak2023real,ni2023beating,cai2024protecting}. Particularly in experiments using bosonic codes, it has been demonstrated that the quality of logical qubits can exceed the so-called break-even point~\cite{sivak2023real,ni2023beating}, validating the effectiveness of quantum error correction techniques in suppressing quantum noise.

Furthermore, to achieve fault-tolerant quantum computing (FTQC), a set of logical gates needs to be implemented. The simplest approach to implement logical gates is transversally, where all physical qubits have interacted with at most one physical qubit from each logical block, therefore naturally ensuring fault-tolerance. However, a well-known theorem states that no quantum code can simultaneously promise a transversal and universal logical gate set~\cite{physrevlett.102.110502,physreva.78.012353,6006592}. For instance, in the surface code, the CNOT gate is transversal. While some single-qubit rotation gates, such as the $S$ gate and $T$ gate, typically need to be implemented indirectly using gate teleportation circuits with ancilla logical states~\cite{PhysRevA.71.022316,PhysRevA.86.032324}.

Currently, more and more experimental works are focusing on demonstrations of logical gates of various quantum error correction codes~\cite{bluvstein2024logical,hu2019quantum,postler2022demonstration,marques2022logical,ryan2022implementing,abobeih2022fault,menendez2023implementing,shaw2024logical,wang2023efficient,hetenyi2024creating}. For instance, in neutral atom systems, demonstrations of the CNOT, CZ, and CCZ gates have been achieved on the [[8,3,2]] color code~\cite{bluvstein2024logical}. In ion trap systems, the $H$, $S$, $T$, and CNOT gates have been demonstrated on the Steane code~\cite{postler2022demonstration}, forming a universal gate set. While in superconducting systems, experimental demonstrations of logical gates are still quite limited. In these systems, the surface code is the most attractive encoding scheme due to its theoretically high threshold and practically nearest-neighbor connectivity requirements~\cite{PhysRevA.86.032324,PhysRevA.83.020302}. Ref.~\cite{marques2022logical} demonstrated a universal set of single-qubit gates on the distance-2 surface code in superconducting systems, showing the potential of using surface code logical qubits for FTQC in the superconducting quantum processor. The main limitation of their work is the lack of two-qubit logical operations, thus not constituting a complete universal gate set. Additionally, the ancilla quantum states used in gate teleportation are physical states rather than logical states, which is inconsistent with the requirements in FTQC. To the best of our knowledge, no work has yet implemented a complete universal set of logical gates in either the superconducting system or the surface code encoding.

\begin{figure}[t]
\begin{center}
\includegraphics[width=0.95\linewidth]{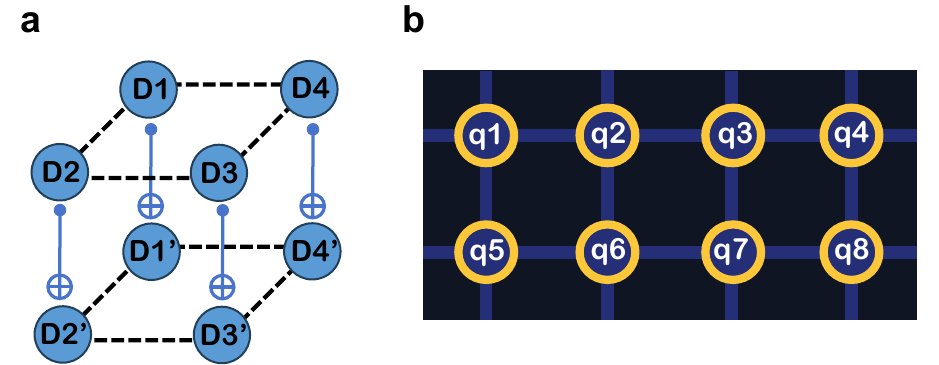}
\end{center}
\setlength{\abovecaptionskip}{0pt}
\caption{\textbf{Distance-2 surface code and qubit layout in the experiment.}
\textbf{(a)} Two logical qubits of the distance-2 surface code and transversal CNOT gate. Each logical qubit is encoded by four data qubits and the logical CNOT gate between two logical qubits corresponds to the four physical CNOT gates between the corresponding data qubits. \textbf{(b)} The experiment uses eight physical qubits arranged in a $2\times4$ rectangular region on the superconducting quantum processor \textit{Wukong}. The deep blue lines represent the topology of the processor, indicating the allowed two-qubit gates between physical qubits.}
\label{fig1}
\end{figure}

In our work, we use the error-detecting surface code with distance 2 (Fig.~\ref{fig1}a) to implement a complete set of universal logical gates, including arbitrary single-qubit rotations around the \( Z \) or \( X \) axis and the CNOT gate, filling the gap in current literature. In the experiment, we encode two logical qubits in a $2\times4$ qubit region of a superconducting quantum processor (see Fig.~\ref{fig1}b and Supplementary Information). The logical CNOT gate is implemented transversally, i.e., by performing four CNOT gates between the corresponding physical qubits. Additionally, single-qubit rotation gates are implemented by preparing the ancilla logical states and applying gate teleportation circuit, which consists of a logical CNOT gate and logical $X$ or $Z$ measurement on the ancilla qubit. {Remarkably, our experiment is not a straightforward extension of previous work but is built upon a tailored design. This design simplifies the encoding of two logical qubits by removing the measurement qubits required for stabilizer measurements, enabling the implementation of transversal CNOT gates on a two-dimensional topology. The error detection in our experiment is achieved through measurement and post-selection at the end of the circuit, without demonstrating stabilizer measurements that involves measurement qubits.} 

The logical Pauli transfer matrices (LPTMs) of these logical gates are characterized on a complete set of states, according to which the gate fidelities are evaluated and listed in Tab.~\ref{table1}. Using fault-tolerant logical state encoding circuits and transversal CNOT gates, four logical Bell states are also prepared. {By verifying the violation of the CHSH inequality with these Bell states, we have confirmed the presence of quantum entanglement between two logical qubits.} In the experiment, all fault-tolerantly prepared logical states, including single-qubit states and Bell states, exhibit higher fidelity than the results on the corresponding physical qubits (see Tab.~\ref{table2}).

{
Note that the fidelity referred to here is the overall fidelity of the preparation and characterization process, therefore it does not indicate that a logical state beyond the break-even point has been achieved. However, this improvement in fidelity suggests that certain specific logical circuits that encoded on detecting codes are expected to perform better than physical circuits in obtaining computational results. With the optimization of hardware, the error rates of these logical operations on error-detecting code are possibly to surpass the break-even point, which could enable more applications in the early FTQC era.}

{Moreover, in the long term, the demonstration of transversal CNOT gates on surface codes could support more efficient FTQC. Theoretical works suggest that combining transversal CNOT gates with two-dimensional (2-D) operations has the potential to reduce the space-time overhead of FTQC on surface codes~\cite{PRXQuantum.4.020345}. However, we recognize that this may be a rather distant goal for superconducting systems, as the transversal CNOT gate for surface codes typically requires a multi-layer architecture or a 2-D architecture with long-distance couplings~\cite{rosenberg20173d,yost2020solid,9134849,gold2021entanglement}. Nonetheless, our experiment provides an early exploration for these intriguing applications.}

\begin{table}[tbp]
    \begin{threeparttable}
    \centering
    \renewcommand{\arraystretch}{1.2}
    \begin{tabular}{c|c|c|c}
        \toprule
        $\,\,$Logical gate $\,\,$ & $\,\,$ Fidelity $\,\,$ & $\,\,$ Logical gate $\,\,$ & $\,\,$ Fidelity $\,\,$\\
        \midrule
        $R_{Z}(0)$ & $94.4(5)\%$ &$R_{X}(0)$ & $92.1(6)\%$  \\
        $R_{Z}(\frac{\pi}{4})$ & $90.0(7)\%$ &$R_{X}(\frac{\pi}{4})$ & $90.7(7)\%$  \\
        $R_{Z}(\frac{\pi}{2})$ & $87.4(7)\%$ &$R_{X}(\frac{\pi}{2})$ &  $89.6(7)\%$  \\
        $R_{Z}({\pi})$ & $93.9(5)\%$ &$R_{X}({\pi})$ & $92.4(6)\%$  \\
        \midrule
        CNOT & $88.9(5)\%$ & &  \\
        \bottomrule
    \end{tabular}
    \end{threeparttable}
    \caption{\textbf{Summary of the fidelities of logical gates in the experiment.}}\label{table1}
\end{table}
\begin{table}[tbp]
\begin{threeparttable}
    \centering
    \renewcommand{\arraystretch}{1.2} 
    \begin{tabular}{c|c|c}
        \toprule
        State & \shortstack{Logical state fidelity } & \shortstack{Physical state fidelity}\\
         \midrule
        $\ket{0_L} / \ket{0}$ & \shortstack{\\$97.9(2)$\%} & $96.9(3)\%$ \\
        $\ket{1_L} / \ket{1}$ & \shortstack{\\$98.0(2)$\%}& $90.8(6)\%$ \\
        $\ket{+_L} / \ket{+}$ & $97.7(2)$\%  & $94.8(4)\%$ \\
        $\ket{-_L} / \ket{-}$ & $97.8(2)$\% & $93.6(5)\%$ \\
        \midrule
          & $79.5(5)\%$& $74.4(9)\%$ \\
          Four Bell& $79.5(5)\%$ & $74.2(9)\%$ \\
          states & $79.4(5)\%$  & $74.5(9)\%$ \\
          & $79.4(5)\%$ & $74.2(9)\%$ \\
        \bottomrule
    \end{tabular}
\end{threeparttable}
\caption{\textbf{Comparison of the fidelities between fault-tolerant prepared logical states and physical states in the experiment.}}\label{table2}
\end{table}

\begin{figure*}[!htbp]
\begin{center}
\includegraphics[width=1\linewidth]{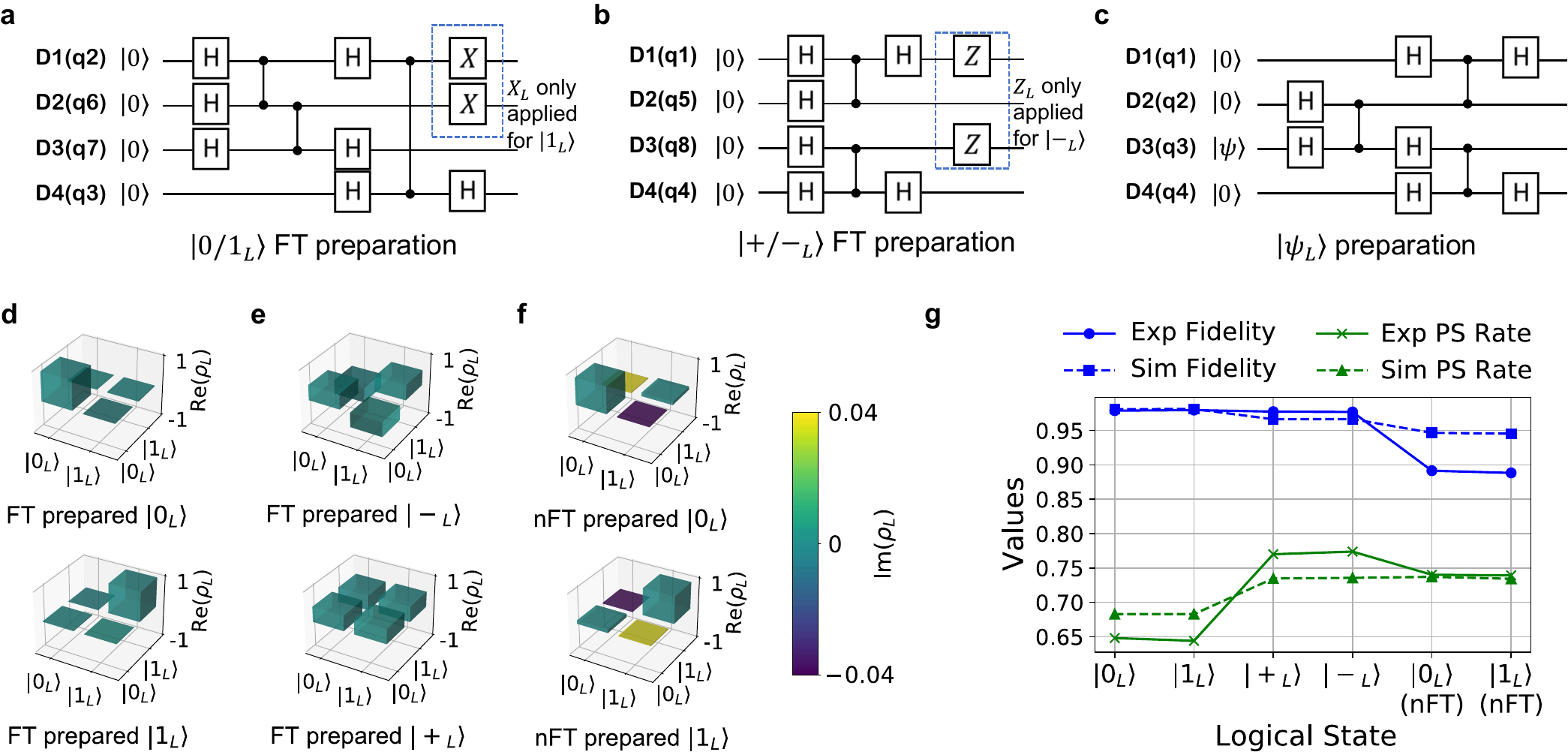}
\end{center}
\setlength{\abovecaptionskip}{0pt}
\caption{\textbf{Logical state preparation circuits and characterization.} 
\textbf{(a-b)} Circuits for fault-tolerant (FT) preparation of $\ket{0/1_L}$ and $\ket{\pm_L}$ states. The $\ket{1_L}$ (or $\ket{-_L}$) state are obtained by applying $X_L$ (or $Z_L$) gate after preparing the $\ket{0_L}$ (or $\ket{+_L}$) state.
\textbf{(c)}
Circuits for non-fault-tolerant (nFT) preparation of arbitrary logical state $\ket{\psi_L}$.
\textbf{(d-f)} Density matrices and fidelities of the six single logical states prepared in the experiment. All logical state density matrices are obtained through logical state tomography.
\textbf{(g)} Comparison of fidelity and post-selection (PS) rates between experiments and simulations. The figure shows the fidelity of six logical states and the post-selection rates when measuring their eigenoperators ($Z_L$ or $X_L$).
}
\label{fig2}
\end{figure*}

\section{Results}
\subsection{Logical state preparation and measurement}
The logical qubit of distance-2 surface code is encoded on four data qubits and is capable of detecting any single-qubit errors. Its code space is the \( +1 \) eigenspace of the following stabilizer group:
\begin{equation}
\mathcal{S} = \langle X_1X_2X_3X_4, Z_1Z_2, Z_3Z_4\rangle .
\end{equation}
Then the logical Pauli operators are defined as:
\begin{equation}
  Z_L = Z_1Z_3, \quad X_L = X_3X_4.
\end{equation}
Accordingly, the explicit form of the logical state can be written as:
\begin{equation}
\begin{aligned}
 |0_L\rangle &= \frac{1}{\sqrt{2}}(|0000\rangle+|1111\rangle), \\
 |1_L\rangle &= \frac{1}{\sqrt{2}}(|0011\rangle+|1100\rangle),
\end{aligned}
\end{equation}
and
\begin{equation}
\begin{aligned}
 |\pm_L\rangle &= \frac{1}{\sqrt{2}}(\ket{0_L}\pm \ket{1_L}).
\end{aligned}
\end{equation}

Here, we designed circuits for preparing the logical states $\ket{0_L}$, $\ket{1_L}$, $\ket{+_L}$ and $\ket{-_L}$ fault-tolerantly (see Fig.~\ref{fig1}), whose fault tolerance is proven in Methods. In order to simultaneously ensure fault-tolerant state preparation and transversal CNOT gate implementation between $\ket{\pm_L}$ and $\ket{0/1_L}$ states, we adopt the qubit allocation scheme depicted in Fig.~\ref{fig2}a and~\ref{fig2}b.  The key is that we exploit the property that $\ket{\pm_L}$ can be decomposed into product states ($\ket{\pm_L}=\frac{1}{2}(\ket{00}\pm\ket{11})^{\otimes2}$), and encode $\ket{\pm_L}$ on the leftmost two (q1 and q5) and the rightmost two physical qubits (q4 and q8) in the hardware. Moreover, we also provide a circuit for preparing arbitrary logical state $\ket{\psi_L}$ in Fig.~\ref{fig2}c. Generally, such a circuit for encoding arbitrary logical state is not fault-tolerant, nor is this circuit. In this way, a logical state can be encoded on a chain of four physical qubits (q1-q4) with only nearest-neighbor coupling. 

After preparing the logical states, logical \(X\), \(Y\), or \(Z\) measurements are performed to characterize these states. Their measurement results are determined by the product of the corresponding Pauli operator measurement result on each data qubits. The logical \(X\) and \(Z\) measurements are fault-tolerant and correspond to measurements in the \(X\) and \(Z\) bases on all data qubits, respectively. Post-selection is carried out based on the conditions provided by the three generators of the stabilizer group, discarding results that violate these conditions. Specifically, assuming the \(X\) or \(Z\) measurement result on the $i$th data qubit is \(m_i^{x}\) or \(m_i^{z} \in \{+1,-1\}\) the post-selection conditions are \(m_1^{x}m_2^{x}m_3^{x}m_4^{x}=+1\), and \(m_1^{z}m_2^{z}=+1\), \(m_3^{z}m_4^{z}=+1\) for logical \(X\) and \(Z\) measurements, respectively. On the other hand, measurement of the logical \(Y\) operator $Y_L=Z_1Y_3X_4$ is not fault-tolerant. It requires  \(Z\) measurements on data qubits D1 and D2, a \(Y\) measurement on D3, and an \(X\) measurement on D4. The corresponding post-selection condition is \(m_1^{z}m_2^{z}=+1\). In this case, post-selection cannot eliminate all single-qubit error cases but can suppress some of them. {Define the probability of successfully passing the post-selection condition as the post-selection rate. Since the post-selection conditions vary under different measurement bases, the post-selection rate is significantly influenced by the measurement basis.}

Here, we conduct experimental demonstrations and characterizations on the fault-tolerantly prepared $\ket{0/1_L}$, $\ket{\pm_L}$ states, and non-fault-tolerantly prepared $\ket{0/1_L}$ states. Through logical quantum state tomography, we constructed the density matrix \(\rho_L\) in the code space, as shown in Fig.~\ref{fig2}~(d-f). Furthermore, we computed the fidelity of the logical state: 
\begin{equation}
   F_L = \langle \psi_L | \rho_L | \psi_L \rangle, 
\end{equation}
where \(|\psi_L\rangle\) is the ideal logical quantum state. The fidelities of the fault-tolerantly prepared states \(|0_L\rangle, |1_L\rangle\) and \(|+_L\rangle, |-_L\rangle\), as well as the non-fault-tolerantly prepared states \(|0_L\rangle\) and \(|1_L\rangle\), are $97.9(2)\%$, $98.0(2)\%$, $97.7(2)\%$, $97.8(2)\%$, $89.2(3)\%$, and $88.9(3)\%$, respectively. We also computed the fidelities of the \(|0\rangle,|1\rangle\) and \(|+\rangle,|-\rangle\) states prepared on the eight physical qubits in the experiment using physical state tomography. For a fair comparison, we did not use readout error mitigation techniques \cite{nation2021scalable} during the physical state tomography. The highest values among eight physical qubits are $96.9(3)\%$ for \(|0\rangle\) in q2, $94.8(4)\%$ for \(|+\rangle\) in q2, $93.6(5)\%$ for \(|-\rangle\) in q2 and $90.8(6)\%$ for \(|1\rangle\) in q3. {All these values are lower than the fidelities of the fault-tolerantly prepared logical states, demonstrating the noise-suppressing effect in the overall process of the preparation and characterization. We remind readers that these fidelities are affected by noise in both the state preparation and the tomography protocol. Due to the difficulty in distinguishing noise in characterization from noise in state preparation, these results do not imply that the fidelity of logical state preparation exceeds that of the physical state. Especially given the significant readout noise on our superconducting processor, the contribution of error detection to the improvement in readout fidelity is likely more substantial.}

{In addition, we provide information on the post-selection rates when measuring the logical state eigenoperators in Fig.~\ref{fig2}e (see Supplementary Information for complete data on the post-selection rate). We also present simulation results for comparison, which are based on the Pauli depolarizing noise model, a commonly used error model in quantum error correction research (also see details in Supplementary Information). However, we also remark that this model does not fully capture the real noise, leading to discrepancies between experimental and simulated data.}

\begin{figure*}[!htbp]
\begin{center}
\includegraphics[width=1.0\linewidth]{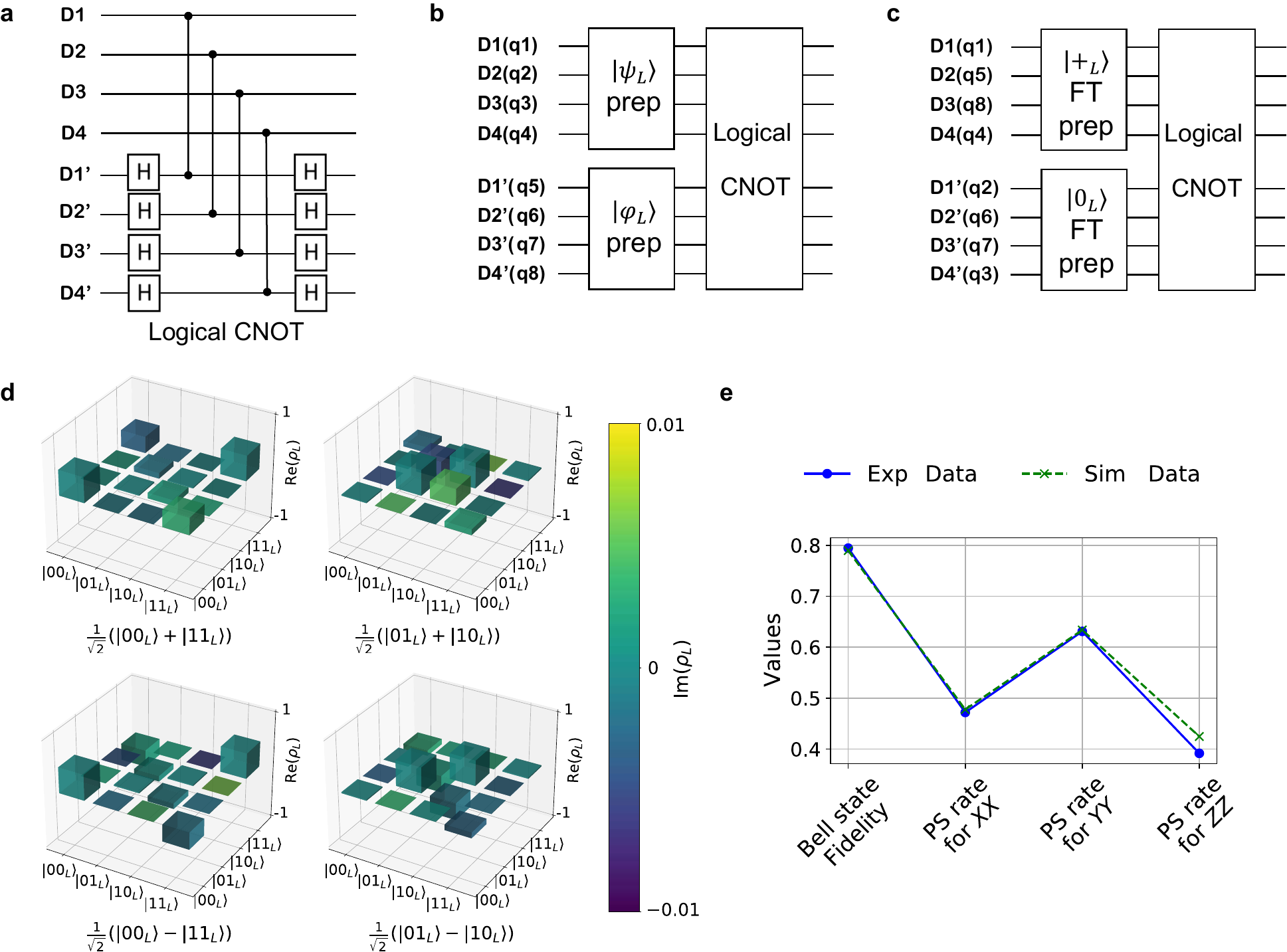}
\end{center}
\setlength{\abovecaptionskip}{0pt}
\caption{\textbf{Logical CNOT gate and Bell state characterization.} 
\textbf{(a)} Circuit of the logical CNOT gate implemented transversally.
\textbf{(b-c)} Circuit for applying a logical CNOT gate on arbitrary logical states $\ket{\psi_L}$ and $\ket{\varphi_L}$, and the circuit for fault-tolerant preparation of Bell states, respectively.  The blocks represent logical state preparation circuits and the logical CNOT gate. The upper half of the logical CNOT block corresponds to the control logical qubit, while the lower half corresponds to the target logical qubit.
\textbf{(d)} Density matrices and fidelities of the four logical Bell states prepared fault-tolerantly in the experiment. 
\textbf{(e)}  Average fidelity and post-selection (PS) rates of four logical Bell states when measuring $X_L\otimes X_L$, $Y_L\otimes Y_L$ and $Z_L\otimes Z_L$ in experiments and simulations.
}
\label{fig3}
\end{figure*}

\subsection{Logical CNOT gate and Bell states}
Next, our experiment demonstrates a transversal CNOT gate between two surface code logical qubits (see Fig.~\ref{fig3}a and \ref{fig3}b). Initially, two logical states $\ket{\psi_L}$ and $\ket{\varphi_L}$, are prepared on two chains of the quantum processor (q1-q4 and q5-q8), where $\ket{\psi_L}$ and $\ket{\varphi_L}$ are from a complete state set $\{\ket{+_L},\ket{-_L},\ket{0_L},\ket{i_L}\}$. Here $\ket{i_L}=(\ket{0_L}+i\ket{1_L})/\sqrt{2}$ is the +1 eigenstate of the logical operator $Y_L$. This step is realized by the preparation circuit for arbitrary logical states described in the previous section. Since the fidelity of states $\ket{+_L}$ and $\ket{-_L}$ in our experiment is higher, we prioritize selecting these two states to form the complete state set. The density matrices of the initial logical states are characterized by logical state tomography. Subsequently, a transversal CNOT gate is applied to the initial logical states, and the output states are characterized using logical state tomography. Based on the expectation values of two-qubit Pauli operators of the initial and output states, we extract the LPTMs
using the method presented in Ref.~\cite{marques2022logical}. The fidelity of the logical CNOT gate, as computed from the LPTM, is found to be $F_L^{G}=88.9(5)\%$. Details concerning the LPTM and fidelity calculation are presented in Methods and Supplementary Information. Due to the noise in the characterization, this result is actually a conservative estimate of the logical gate fidelity.

Then we use the logical CNOT gate to prepare four Bell states on logical qubits, which are important entangled resources in quantum information. Following the above initialization method, the control and target logical qubits can be initialized to $\ket{\pm_L}$ and $\ket{0/1_L}$  states, respectively. Then they can be acted by a logical CNOT gate to generate a Bell state. However, under such qubit allocation, the prepared $\ket{0/1_L}$ state is not fault-tolerant. Therefore, we adopt the qubit allocation scheme from the previous section to simultaneously fault-tolerantly prepare the $\ket{0/1_L}$ and $\ket{\pm_L}$ states (see Fig.~\ref{fig3}c). This circuit can be viewed as a special planarization of a two-layer architecture. In this layout, all physical CZ gates required in both the logical state preparation and the transversal CNOT gate implementation are 2-D hardware-neighbor. We reconstruct the density matrix of the logical Bell states in Fig.~\ref{fig3}d. {The overall fidelities in the preparation and characterization for the four logical Bell states are $79.5(5)\%$, $79.5(5)\%$, $79.4(5)\%$, and $79.4(5)\%$, respectively. We also report the post-selection rates for Bell states under $X\otimes X$, $X\otimes X$, $Z\otimes Z$ measurements along with a comparison between simulated and experimental data in Fig.~\ref{fig3}e.} Correspondingly, we prepare four physical Bell states by physical CNOT gate on qubits q6 and q7. The fidelity of the CNOT gate between q6 and q7 is the highest among all physical CNOT gates in the experiment. The fidelities for the four physical Bell states are $74.4(9)\%$, $74.2(9)\%$, $74.5(9)\%$, and $74.2(9)\%$ respectively, all of which are lower than the fidelity of the fault-tolerantly prepared logical Bell states.

{To confirm entanglement between the two surface code logical qubits, we verify a variant of the CHSH inequality~\cite{HORODECKI1995340}. For a two-qubit density matrix $\rho$, define the matrix $T_\rho$ with elements $(T_\rho)_{ij} = \text{Tr}(\rho P_i \otimes P_j )$, where $P_i \in \{X, Y, Z\}$. A necessary and sufficient condition for violating the CHSH inequality is $u_1 + u_2 > 1$, where $u_1$ and $u_2$ are the two largest eigenvalues of the matrix $T_\rho^T T_\rho$. In our experiment, the values of $u_1 + u_2$ for the four logical Bell states are 1.55, 1.55, 1.54, and 1.54, respectively. This result confirms the presence of quantum entanglement between the two surface code logical qubits.}

\subsection{Logical single-qubit rotation}

\begin{figure*}[!htbp]
\begin{center}
\includegraphics[width=1.0\linewidth]{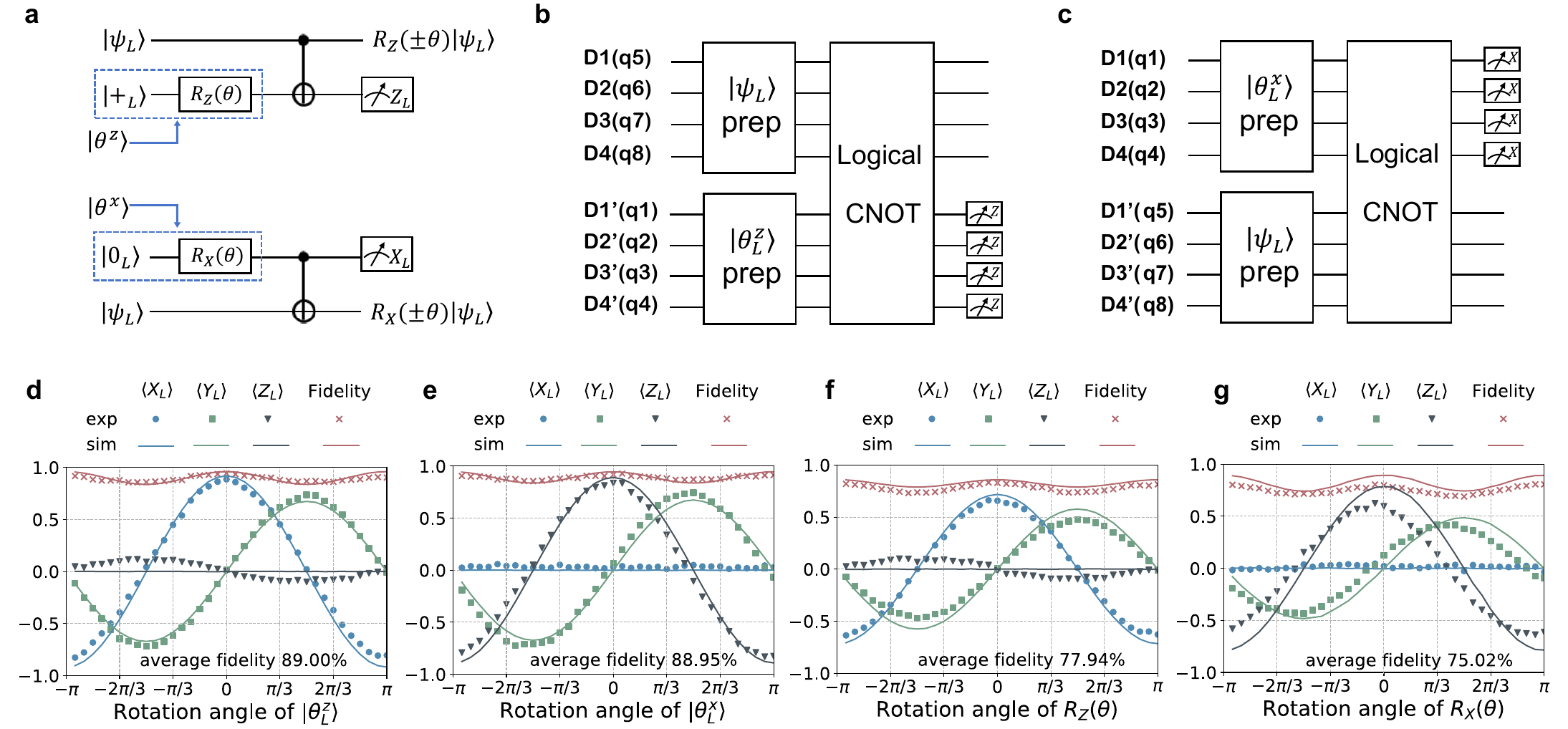}
\end{center}
\setlength{\abovecaptionskip}{0pt}
\caption{\textbf{Logical single-qubit rotations and characterization.} \textbf{(a)} Gate teleportation circuits that implement single-qubit rotation operations on logical qubits. The $\pm$ sign of the rotation angle depends on the measurement results of the ancilla logical states.
\textbf{(b-c)} Circuits for applying single-qubit rotations \( R_Z (\theta) \) and \( R_X (\theta) \) on the logical state $\ket{\psi_L}$ based on gate teleportation, respectively.
\textbf{(d-e)} Average values of Pauli operators and fidelity of the ancilla logical states $|\theta^z_L \rangle$ and $|\theta^x_L \rangle$ with rotation angles \( \theta \in (-\pi, \pi] \), respectively. Scatter points and solid lines are used to distinguish experimental and simulated data.
\textbf{(f-g)} Average values of Pauli operators and fidelity of the output states $R_Z (\theta)|+_L \rangle$ or $R_X (\theta)|0_L \rangle$ with rotation angles \( \theta \in (-\pi, \pi] \), respectively.
}
\label{fig4}
\end{figure*}

Finally, we demonstrated logical single-qubit rotations around the \( Z \) or \( X \) axis based on gate teleportation circuit (Fig.~\ref{fig4}a). More specifically, these rotation operations are 
\begin{equation}
\begin{aligned}
 R_Z (\theta)=e^{-i\theta Z_L/2}, \quad
 R_X (\theta)=e^{-i\theta X_L/2},
\end{aligned}
\end{equation}
where $\theta$ is the rotation angle. The gate teleportation circuit consists of three parts. First, preparing the ancilla states 
 \begin{equation}
\begin{aligned}
 |\theta^z_L \rangle &= \frac{1}{\sqrt{2}}(|0_L\rangle + e^{i\theta} |1_L\rangle),\\
 |\theta^x_L \rangle &= \cos \frac{\theta}{2} |0_L\rangle - i\sin \frac{\theta}{2} |1_L\rangle. 
 \end{aligned}
\end{equation}
 Then the logical CNOT gate is applied, and finally, ancilla state is measured in logical \( Z \) or \( X \) basis. The \( R_Z (\theta) \) or \( R_X (\theta) \) gate is successfully executed only when the logical \( Z \) or \( X \) measurement results in $+1$; otherwise, operation \( R_Z (2\theta) \) or \( R_X (2\theta) \) needs to be applied as a compensation. Here, we simply use the post-selection strategy, that is, only retaining the cases where the measurement result is $+1$. Note that the ancilla states can be viewed as the result of applying $R_Z (\theta)$ or $R_X (\theta)$ gates to $|+_L\rangle$ or $|0_L\rangle$, respectively, that is why we refer to this circuit as gate teleportation circuit. 

In the experiment, we first prepare the required ancilla logical states $|\theta^z_L \rangle$ and $|\theta^x_L \rangle$ with \( \theta \in (-\pi, \pi] \) on a chain of the quantum processor (q1-q4). Then these input states are measured in $X_L$, $Y_L$ or $Z_L$ basis to obtain the expectation values of the logical Pauli operators.  Subsequently, we execute the circuits in Fig.~\ref{fig4}b and \ref{fig4}c, demonstrating the single-qubit rotation gates around the \( Z \) or \( X \) axis on the state \( |\psi_L\rangle=|+_L\rangle \) or \( |0_L \rangle \), respectively. The expectation values of the logical Pauli operators for the input and output states are shown in Fig.~\ref{fig4}~(d-g). Using the expectation values $\langle X\rangle$, $\langle Y\rangle$, $\langle Z\rangle$, we reconstructed the density matrices, thereby calculating the fidelity of each state. The average fidelities of input states \( |\theta^z_L \rangle \) and \( |\theta^x_L \rangle \) are evaluated to be $89.0(3)\%$. Correspondingly, the average fidelities of the output states are $78.0(9)\%$ and $75.0(9)\%$, respectively. 

To characterize the fidelity of the single-qubit logical gates, it is required to construct the LPTMs of these gates. Here, we test the LPTMs of \( R_Z(\theta) \) and \( R_X(\theta) \) with \( \theta\in\{0,\pi/4,\pi/2,\pi\}\) as examples. The input states are encoded as the logical states from the set \( \{ \ket{+_L}, \ket{-_L}, \ket{0_L}, \ket{i_L} \} \), and the above logical gates are applied separately. We measure the expectation values of the Pauli operators for the input and output states and construct the LPTMs for these eight logical gates accordingly (see Methods and Supplementary Information). The fidelities $F_L^G$ of these eight logical gates are estimated to be $94.4(5)\%$, $90.0(7)\%$, $87.4(7)\%$, $93.9(5)\%$, and $92.1(6)\%$, $90.7(7)\%$, $89.6(7)\%$, $92.4(6)\%$, respectively.

\section{discussion}

{This work experimentally demonstrates a complete universal set of logical gates on distance-2 surface code in a superconducting processor. Particularly, logical Bell states that violates CHSH inequality have been fault-tolerantly prepared using the transversal CNOT gate. Based on the logical CNOT gate, the gate teleportation process is experimentally demonstrated to implement single-qubit rotation operations. These results reveal several significant aspects of FTQC based on the surface code in superconducting hardware.}

{The fidelity of logical operations are in the experiment is affected by a variety of factors. The dominant noise of our superconducting processor is the readout noise and two-qubit gate noise. Through numerical simulations, we found that the performance of logical circuits in our experiment is more sensitive to readout errors compared to gate errors. The Supplementary Information present the results of these numerical simulations and discuss the mechanisms underlying various types of noise as well as potential approaches for improvement.} In addition, in the implementation of single-qubit rotation gates, the fidelity of the logical gates largely depends on the quality of the ancilla logical states in the gate teleportation circuit. In our experiment, the ancilla logical states are generated by non-fault-tolerant preparation circuits, resulting in a relatively high error rate. In a complete FTQC framework, high-fidelity ancilla logical states are typically obtained through state distillation~\cite{PhysRevA.71.022316,PhysRevA.86.052329,Litinski2019magicstate,Campbell_2016}. A particularly challenging future task is to experimentally demonstrate these distillation protocols.

{In our experiment, logical qubits are confined to a one-dimensional structure without measurement qubits. A natural extension is to incorporate the repeated stabilizer measurement process into our work. Achieving both the stabilizer measurement process and transversal CNOT gate typically requires a multi-layer structure or long-range entangling gates (see Supplementary Information). For superconducting platforms, this is regarded as a challenging long-term goal. However, we are also excited to see that they are increasingly gaining attention due to the requirements in FTQC~\cite{10.1063/5.0082975,bravyi2024high, ramette2024fault}. Meanwhile, some prototypes of these technologies have been demonstrated recently~\cite{rosenberg20173d,yost2020solid,9134849,gold2021entanglement}, indicating that they are not beyond reach.}

{In conclusion, our experiment enriches the possibilities for research in FTQC. First, from a near-term perspective, our work demonstrates the role of error-detecting codes or small-distance error-correction codes in the early FTQC era. Notably, the performance of some logical circuits in the experiment surpassed that of physical circuits, suggesting that applying error detection may benefit certain small-scale quantum tasks. Numerical simulations further indicate that the pseudo-threshold of the experimental circuits can significantly exceed the fault-tolerant threshold (approximately 1\%), making them more practical for applications in the early FTQC era (see Supplementary Information). Second, on superconducting platforms with planar nearest-neighbor connectivity, lattice surgery is the mainstream method for logical operations~\cite{Horsman_2012,Litinski2018latticesurgery,Litinski2019gameofsurfacecodes}. Demonstrating transversal CNOT gates supports a hybrid approach combining them with lattice surgery, potentially reducing the significant overhead of FTQC~\cite{PRXQuantum.4.020345, viszlai2023architecture}. Achieving this requires extending the experimental qubit layout to a multi-layer structure, which remains a long-term goal for superconducting platforms.}

\section{methods}

\subsection{Fault-tolerant logical state preparation}
\begin{figure}[t]
\begin{center}
\includegraphics[width=1\linewidth]{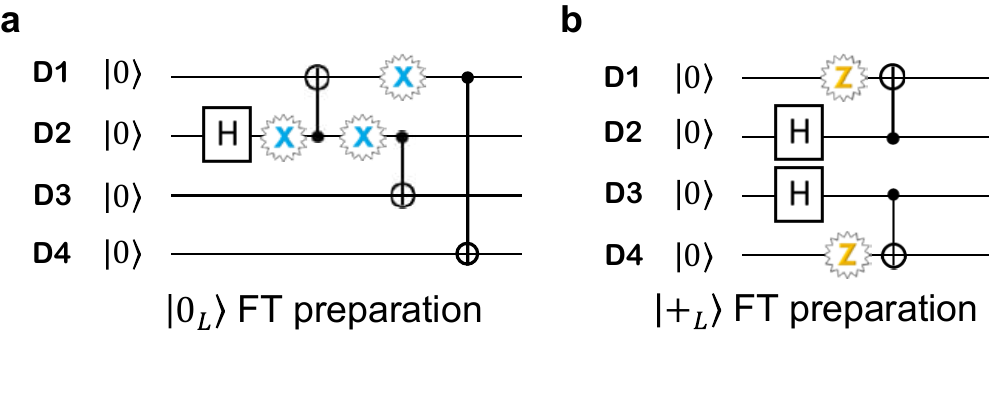}
\end{center}
\setlength{\abovecaptionskip}{0pt}
\caption{\textbf{Equivalent logical state fault-tolerant preparation circuit.} The circuits are simplified to a composition of CNOT and $H$ gates, with fault tolerance equivalent to the original circuits. The possible $X$ (blue) or $Z$ (yellow) errors that could propagate are shown.
\textbf{(a)} Fault-tolerant (FT) preparation circuit for $|0\rangle$.
\textbf{(b)} Fault-tolerant (FT) preparation circuit for $|+\rangle$.}
\label{fig5}
\end{figure}

Here we prove that the circuits in the first two parts of Fig.~\ref{fig2}a and \ref{fig2}b are fault-tolerant, meaning that a single-qubit error occurring at any position in the circuit can be detected without leading to a logical error. To clarify this, we note that there are two types of errors to consider: those that remain localized in a single qubit and are thus detectable by the stabilizers, and those that might affect the final state of more than one qubit. We focus on the latter type of errors, ensuring that they do not spread to become logical errors. For ease of discussion, we combine the \( H \) gates and CZ gates in the circuit into CNOT gates, focusing on the preparation of the \( \ket{0_L} \) and \( \ket{+_L} \) states, resulting in the circuit shown in Fig.~\ref{fig5}. This simplification does not affect the fault-tolerance of the original circuits.

For the \( \ket{0_L} \) state preparation circuit, we only need to consider the Pauli \( X \) errors in the circuit, as any logical \( Z_L \) error produced is trivial for the \( \ket{0/1_L} \) state up to a global phase. We mark the locations of all possible single-qubit Pauli \( X \) errors (shown as blue \( X \) in Fig.~\ref{fig5}a). The leftmost \( X \) error affects qubits 1 through 4 as \( X_1X_2X_3X_4 \), which is a stabilizer. The second and third \( X \) errors affect qubits 2 and 3 as \( X_2X_3 \) and qubits 1 and 4 as \( X_1X_4 \), respectively. These errors anti-commute with the stabilizers \( Z_1Z_2 \) and \( Z_3Z_4 \), and thus they will be detected by the stabilizer measurements. This proves that no single-qubit Pauli \( X \) error at any position in the circuit can spread to become a logical \( X_L \) error.

Similarly, in the \( \ket{+_L} \) state preparation circuit, we consider the possible Pauli \( Z \) errors. The two possible spreading Pauli \( Z \) errors (yellow \( Z \) in Fig.~\ref{fig5}b) affect qubits 1 and 2 as \( Z_1Z_2 \) and qubits 3 and 4 as \( Z_3Z_4 \), which are the two stabilizers of this code. Since all these errors can be detected or lead to a stabilizer operator, we have demonstrated the fault-tolerance of these two encoding circuits.

\subsection{Logical Pauli transfer matrix (LPTM)}
The Pauli transfer matrix (PTM) describes a quantum process on the components of the density matrix represented in the basis of Pauli operators~\cite{nielsen2010quantum,PhysRevLett.109.060501,greenbaum2015introduction,Nielsen2021gatesettomography}. For a \( d \)-dimensional Hilbert space, a PTM \( \mathcal{R} \) is a linear transformation matrix from the expectation values \( p_i = \langle P_i \rangle \) of the Pauli operators \( P_i \) in the input state to the expectation values \( p'_j \) in the output state:
\begin{equation}
  p'_j = \sum_i \mathcal{R}_{ij} p_i.  
\end{equation}
In our experiment, \( P_i \) belongs to \( \{I_L,X_L,Y_L,Z_L\}^{\otimes 2}\) and \( \{I_L,X_L,Y_L,Z_L\}\) for the cases \( d = 4 \) and \( d = 2 \), respectively. To construct the LPTMs of the logical quantum gates in the main text, we use input states from the complete set \( \{ \ket{+_L}, \ket{-_L}, \ket{0_L}, \ket{i_L} \}^{\otimes 2}\) (for the logical CNOT gate) or \( \{ \ket{+_L}, \ket{-_L}, \ket{0_L}, \ket{i_L} \} \) (for the logical single-qubit gates). The density matrices of the input and output states are obtained through logical state tomography, and the expectation values \( p_i \) and \( p'_j \) are then calculated. The inverse of the expectation value matrix yields the raw result \( \mathcal{R}^{\text{raw}} \). However, \( \mathcal{R}^{\text{raw}} \) may not satisfy the conditions of a physical channel, i.e., being completely positive and trace-preserving~\cite{nielsen2010quantum}. Therefore, using the techniques in Ref.~\cite{marques2022logical}, \( \mathcal{R}^{\text{raw}} \) is transformed into the Choi state representation:
\begin{equation}
    \rho_{\text{choi}} = \frac{1}{d^2} \sum_{ij} \mathcal{R}^{\text{raw}}_{ij} P^T_j \otimes P_i.
\end{equation}
We then optimize \( \rho \) under the following objective function and constraints:
\begin{equation}
\begin{aligned}
\text{minimize} \quad & \sum_{i,j} \left| \text{Tr}(\rho P^T_j \otimes P_i) - \mathcal{R}^{\text{raw}}_{ij} \right|^2, \\
\text{subject to} \quad & \rho \geq 0, \text{Tr}(\rho) = 1, \text{Tr}_1(\rho) = \frac{1}{2}\mathbbm{1},
\end{aligned}
\end{equation}
where $\text{Tr}_1$ is the partial trace over the left half subsystem. Using the convex optimization package \textit{cvxpy}, we obtain the optimal result \( \rho_{\text{opt}} \). The corresponding LPTM \(\mathcal{R}\) is
\begin{equation}
\mathcal{R}_{ij} = \text{Tr}(\rho_{\text{opt}} P^T_j \otimes P_i)
\end{equation}
and the fidelity of the logical gate is
\begin{equation}
F_L^G =  \frac{\text{Tr}(\mathcal{R}^\dag \mathcal{R}_{\text{ideal}}) +d}{d^2 + d},
\end{equation}
where \(\mathcal{R}_{\text{ideal}} \) is the ideal LPTM of the logical gate. In our experiment, we constructed the LPTMs for the logical CNOT gate and eight logical single-qubit gates. The specific details of these LPTMs can be found in Supplementary Material Fig.~\ref{figs2} and Fig.~\ref{figs3}.

\subsection{Quantum state tomography}
Quantum state tomography~\cite{PhysRevLett.72.1137,PhysRevLett.74.4101,cramer2010efficient} reconstructs the density matrix of an unknown quantum state by measuring some observables. In our experiment, we measure \(4^n-1\) Pauli operators of the logical qubits, where \(n\) is the number of logical qubits. Assuming the expectation values of these Pauli operators are \( p_i = \langle P_i \rangle \), where \( P_i \in  \{I_L,X_L,Y_L,Z_L\}^{\otimes n}  / \{I_L^{\otimes n} \} \), the density matrix is reconstructed as:
\[
\rho_{L,0} = \sum_{i=0}^{4^n-1} \frac{p_i P_i}{2^n},
\]
with \( p_0 = 1 \) and \( P_0 = I_L^{\otimes n} \). Such a density matrix $\rho_{L,0}$ may not satisfy the physicality characteristics of a quantum state. Therefore, we use maximum likelihood estimation~\cite{PhysRevA.61.010304,PhysRevLett.108.070502} to construct the logical density matrix \( \rho_L \). Specifically, the objective function to minimize is
\begin{equation}
   \sum_{i} | \text{Tr}(\rho_L P_i) - p_{i} |^2, 
\end{equation}
subject to \( \text{Tr}(\rho_L) = 1 \), and \( \rho_L \geq 0 \). This process is implemented also using the convex optimization package \textit{cvxpy}. Likewise, we also apply state tomography to physical states for constructing the density operators of states $\ket{0}$, $\ket{1}$, $\ket{+}$, $\ket{-}$ and four Bell states, which is done for comparison with the logical state density matrices. These results are shown in Supplementary Material Fig.~\ref{figs4}.

{
\section{Data availability}
The data that support the findings of this study are available from the corresponding author upon reasonable request.}

\section{acknowledgments}
We thank Prof.~Chang-Ling Zou and Prof.~Ying Li for reviewing the manuscript and providing valuable suggestions, and thank Cheng Xue and Xi-Ning Zhuang for their assistance reviewing this manuscript. This work is supported by National Key Research and Development Program of China (Grant No. 2023YFB4502500)

\bibliographystyle{apsrev4-2}
\bibliography{ref}

\clearpage
\setcounter{table}{0}
\renewcommand{\thetable}{S\arabic{table}}%
\setcounter{figure}{0}
\renewcommand{\thefigure}{S\arabic{figure}}%
\setcounter{section}{0}
\setcounter{equation}{0}
\renewcommand{\theequation}{S\arabic{equation}}%

\onecolumngrid
\begin{center}
{\large \bf Supplementary Information}
\vspace{0.3cm}
\end{center}

\setcounter{page}{1}

\section{device}
\begin{figure}[htbp]
\begin{center}
\includegraphics[width=0.5\linewidth]{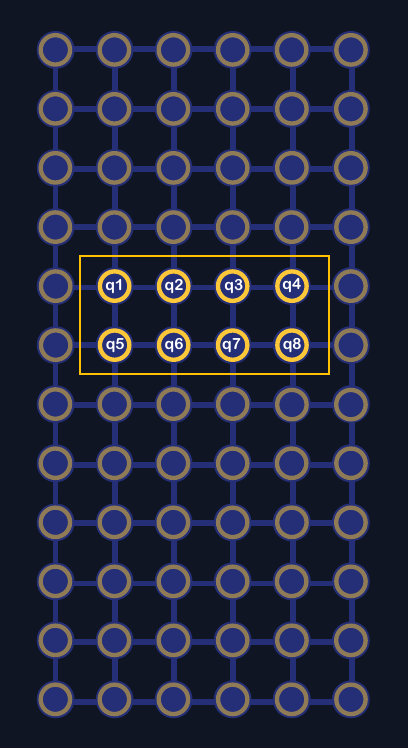}
\end{center}
\setlength{\abovecaptionskip}{0pt}
\caption{\textbf{The topology of the superconducting quantum processor \textit{Wukong} and the region used in the experiment.}}
\label{figs1}
\end{figure}

The superconducting quantum processor \textit{Wukong} consists of 72 transmon superconducting qubits (62 available) arranged with a square lattice topology. The basic physical gate set on this processor includes \( I \), \( X \), \( \sqrt{X} \), \( R_z(\theta) \), and \( CZ \), where \( R_z(\theta) \) is implemented via virtual \( Z \) technique~\cite{PhysRevA.96.022330}. The execution time for \( I \) (or \( R_z(\theta) \)), \( X \) (or \( \sqrt{X} \)) and \( CZ \) gate are 0 ns, 30 ns and 40 ns, respectively. Note that by decomposing 
$H$$=$\( R_z(\frac{\pi}{2}) \)\( \sqrt{X} \)\( R_z(\frac{\pi}{2}) \), the Hadamard gate can be executed within a single gate time (30 ns). In the experiment, we use a $2\times4$ qubit rectangular region whose positions are shown in Fig.~\ref{figs1}. The single-qubit parameters and fidelities of CZ gates in the region are presented in Tab.~\ref{tables1} and Tab.~\ref{tables2}, respectively.

{All circuits in the experiment are submitted by a cloud platform and sampled over $5\times10^4$ times in total, including the sum of counts that passed and failed the post-selection.} We disabled circuit optimization and use barriers to fix the time order of the gates. In all circuits, the $H$-layer and CZ-layer are separated to avoid $XY$ crosstalk~\cite{barends2014superconducting}. When preparing two logical states simultaneously, we stagger the timing of one $H$-layer and one CZ-layer to avoid potential CZ crosstalk~\cite{PhysRevApplied.14.024042}. Additionally, the transversal CNOT gate is executed in two CZ layers, 
ensuring that CZ gates with high crosstalk do not execute in parallel.

{In our superconducting quantum processor, noise sources for single- and two-qubit gates are categorized as isolated and parallel errors. For single-qubit gates, isolated noise sources primarily include decoherence due to qubit relaxation and dephasing, and waveform control imperfections leading to unitary deviations. Parallel noise sources involve microwave crosstalk, where signals from adjacent qubits interfere, and residual ZZ coupling, an incomplete coupling cancellation between qubits. Decoherence and microwave crosstalk currently represent the primary factors that limit gate fidelity for single-qubit operations.}

{Two-qubit gate errors arise from similar isolated and parallel sources. Isolated errors involve decoherence from two-level system interactions and leakage due to imperfect pulse control, while waveform distortions lead to operational inaccuracies. Parallel errors primarily stem from frequency collisions, causing unintended spurious couplings between qubits operating at overlapping frequencies. In two-qubit gates, decoherence and leakage from control imperfections limit fidelity. Improvements focus on refining chip fabrication for coherence, optimizing algorithmic methods for frequency allocation, and enhancing pulse design with techniques such as reinforcement learning to reduce control-induced errors.}

{Moreover, in our superconducting processors, readout errors mainly arise from assignment (overlap) error, thermal excitation error, preparation error, relaxation error, and leakage error. Assignment error occurs when there is incomplete separation between the 0 and 1 signals, leading to misclassification. To reduce this, a higher signal-to-noise ratio (SNR) is needed, achievable with quantum-limited noise amplifiers. Thermal excitation error, caused by an initial thermal population of the qubit in an excited state, can be minimized through improved chip packaging for better thermal conduction and dissipation, or by using post-selection and heralding techniques. Preparation error, resulting from imperfections in gate operations, has a generally smaller impact on readout fidelity and can be further reduced by enhancing gate calibration and control. Relaxation error arises when the qubit relaxes from the excited to the ground state during readout, primarily affecting the fidelity of excited-state measurements. Shortening the readout duration can mitigate this, though it requires a higher SNR. Finally, leakage error occurs when the qubit is excited to states beyond the computational basis due to excessive readout amplitude; choosing suitable readout pulse amplitudes can prevent this, especially in mid-circuit measurements such as those used in quantum error correction. Together, these errors impact the accuracy of readout processes.}

\begin{table}[h]
\centering
\begin{tabular}{lccccc}
\toprule
Qubit / parameter & $\quad T_1 (\mu s)\quad$ & $\quad T_2 (\mu s)\quad$ & Readout $F_{00} (\%)$ & Readout $F_{11} (\%)$ & Single-qubit gate error $(\%)$ \\
\midrule
q1  & 29.3569 & 1.215401 & 96.10 & 91.50 & 0.24 \\
q2  & 14.4954 & 1.740412 & 97.18 & 87.12 & 0.25 \\
q3  & 10.0308 & 1.854965 & 97.80 & 90.38 & 0.18 \\
q4  & 13.1754 & 2.287143 & 96.28  & 92.28 & 0.40  \\
q5  & 12.2130 & 2.792981 & 93.86 & 88.72 & 0.25 \\
q6  & 17.3159 & 3.009932 & 93.36 & 89.28 & 0.19 \\
q7  & 13.6927 & 2.057180 & 97.00 & 86.16 & 0.22 \\
q8  & 21.9188 & 3.115603 & 91.46 & 87.36 & 0.25 \\
\midrule
Average & 16.5249 & 2.259202 & 95.38 & 89.10 & 0.25 \\
\bottomrule
\end{tabular}
\caption{Summary of single-qubit parameters}
\label{tables1}
\end{table}

\begin{table}[h]
\centering
\begin{tabular}{p{4cm}|c|c|c|c|c|c|c|c|c|c|c}
\toprule
Qubit pair &  q1-q2  & q2-q3 & q3-q4 & q5-q6 & q6-q7 & q7-q8 & q1-q5 & q2-q6 & q3-q7 & q4-q8 & Average \\
\midrule
CZ fidelity (XEB) (\%) & 95.55 & 96.18 & 98.39 & 94.18 & 98.64 & 97.09 & 97.89 & 97.45 & 96.42 & 98.44 & 97.02 \\
\bottomrule
\end{tabular}
\caption{Summary of CZ fidelity (XEB)}
\label{tables2}
\end{table}

\clearpage
\section{LPTMs of logical gates and physical state characterization}
\begin{figure*}[!htbp]
\begin{center}
\includegraphics[width=0.75\linewidth]{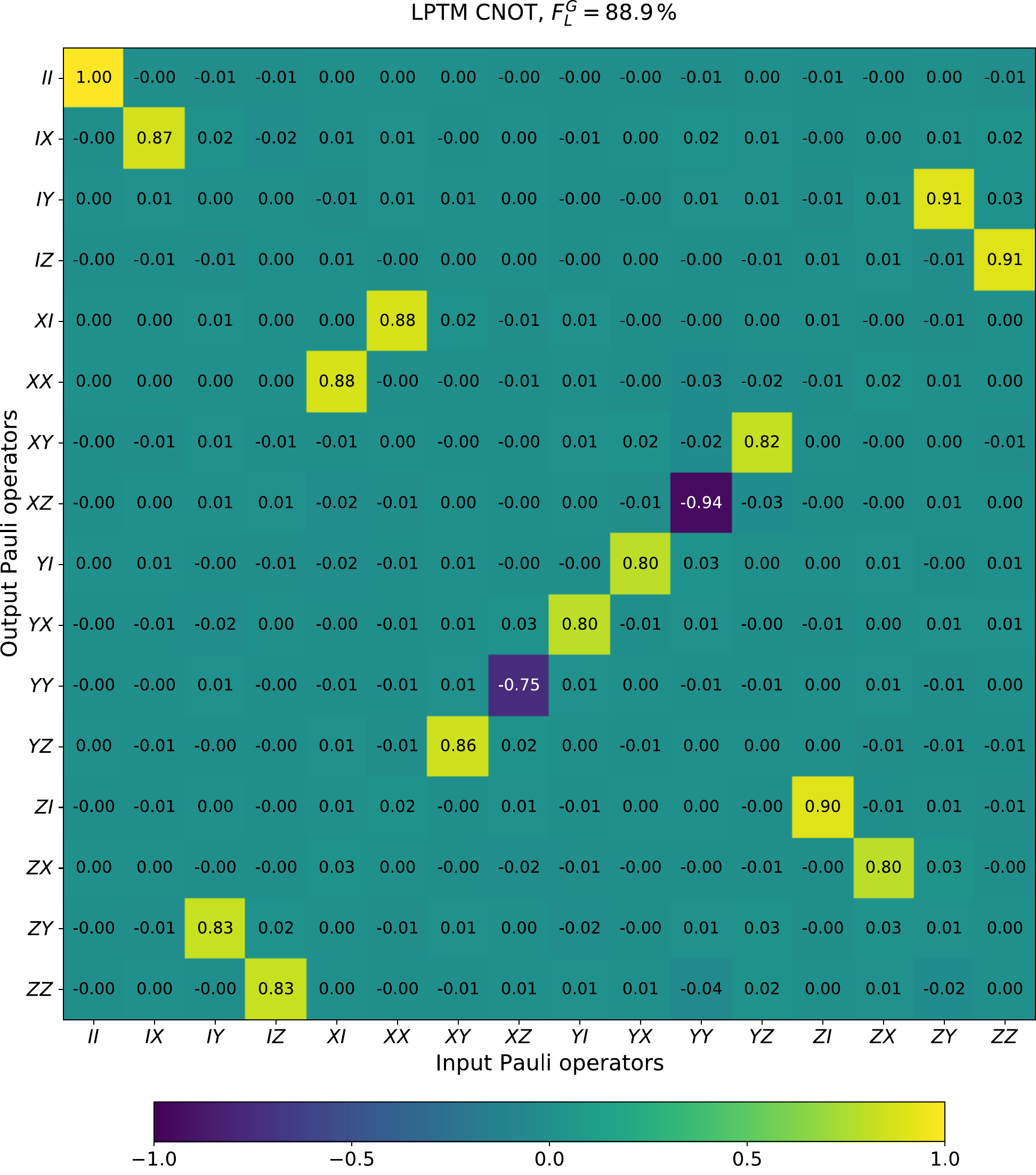}
\end{center}
\setlength{\abovecaptionskip}{0pt}
\caption{\textbf{Characterization of the logical CNOT gate in the experiment by the logical Pauli transfer matrix.}}
\label{figs2}
\end{figure*}

\begin{figure*}[t]
\begin{center}
\includegraphics[width=0.8\linewidth]{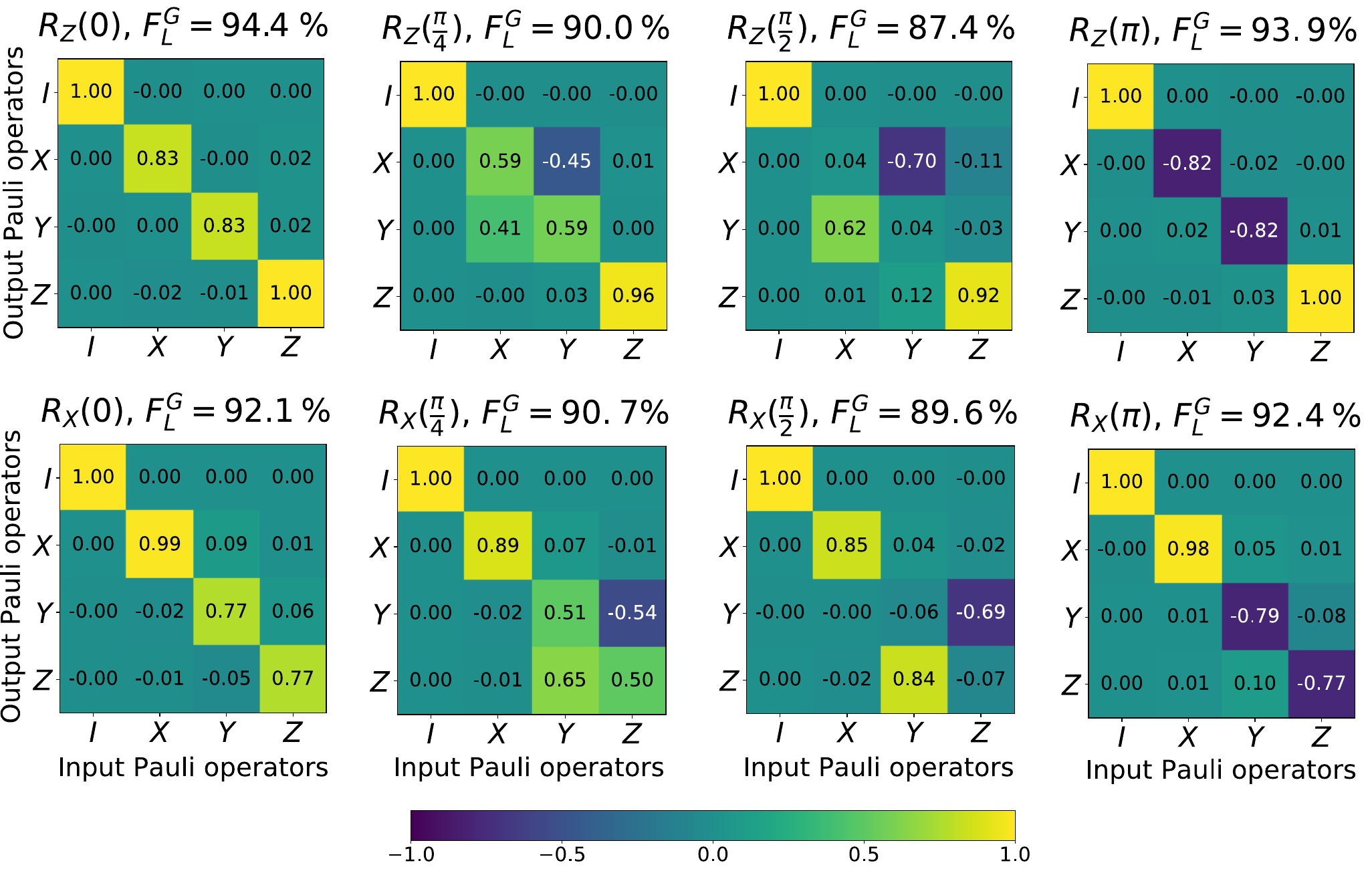}
\end{center}
\setlength{\abovecaptionskip}{0pt}
\caption{\textbf{Characterization of the logical logical single-qubit rotation gates in the experiment by the logical Pauli transfer matrices.}}
\label{figs3}
\end{figure*}

\begin{figure*}[t]
\begin{center}
\includegraphics[width=0.95\linewidth]{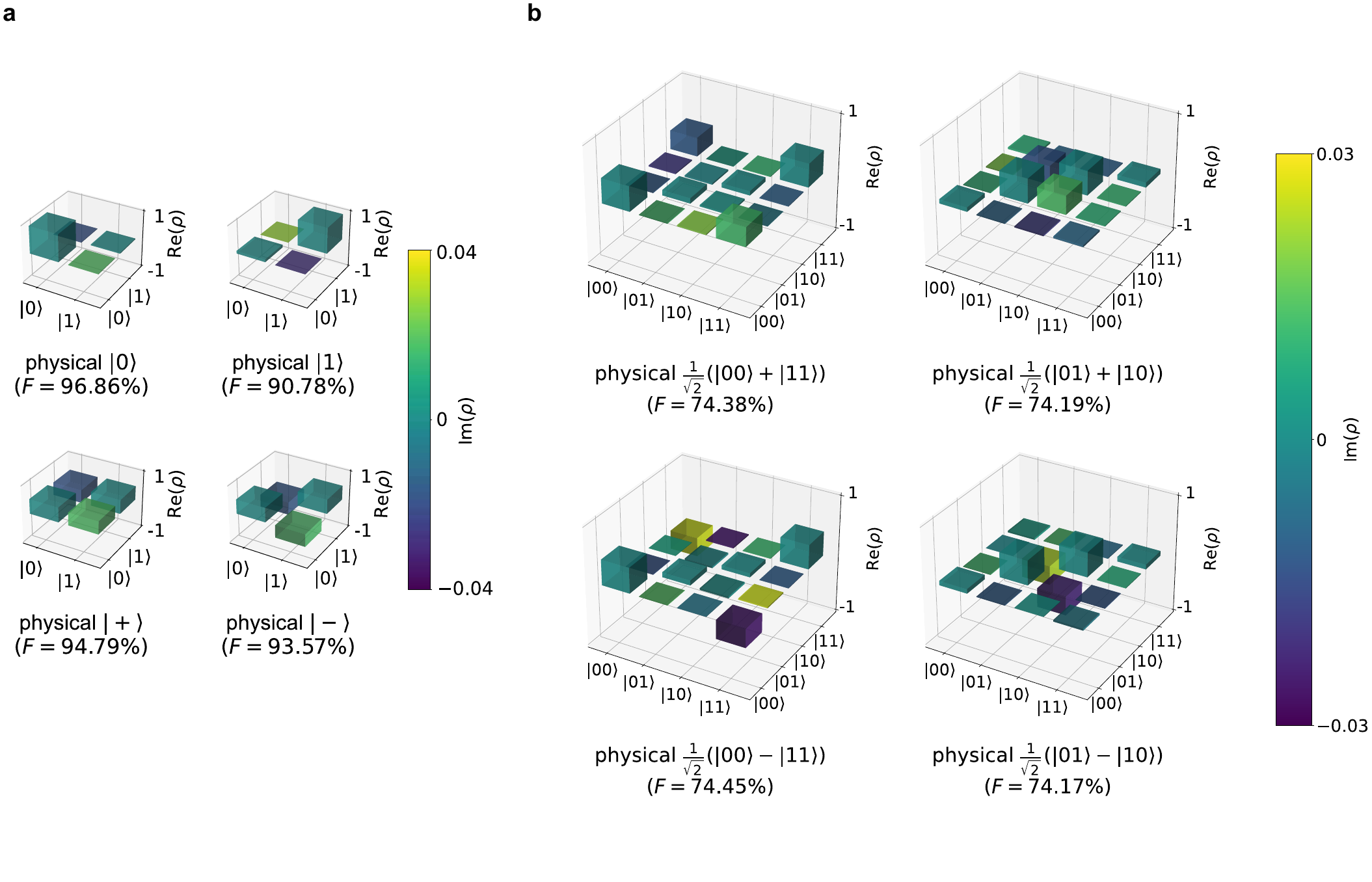}
\end{center}
\setlength{\abovecaptionskip}{0pt}
\caption{\textbf{Characterization of the physical states.} 
\textbf{(a)} Density matrices and fidelities of the physical states $\ket{0}$, $\ket{1}$, $\ket{+}$ and $\ket{-}$.
\textbf{(b)} Density matrices and fidelities of the physical Bell states.}
\label{figs4}
\end{figure*}

\clearpage
{
\section{Data of post-selection rates}}

{\begin{figure*}[htbp]
\begin{center}
\includegraphics[width=0.95\linewidth]{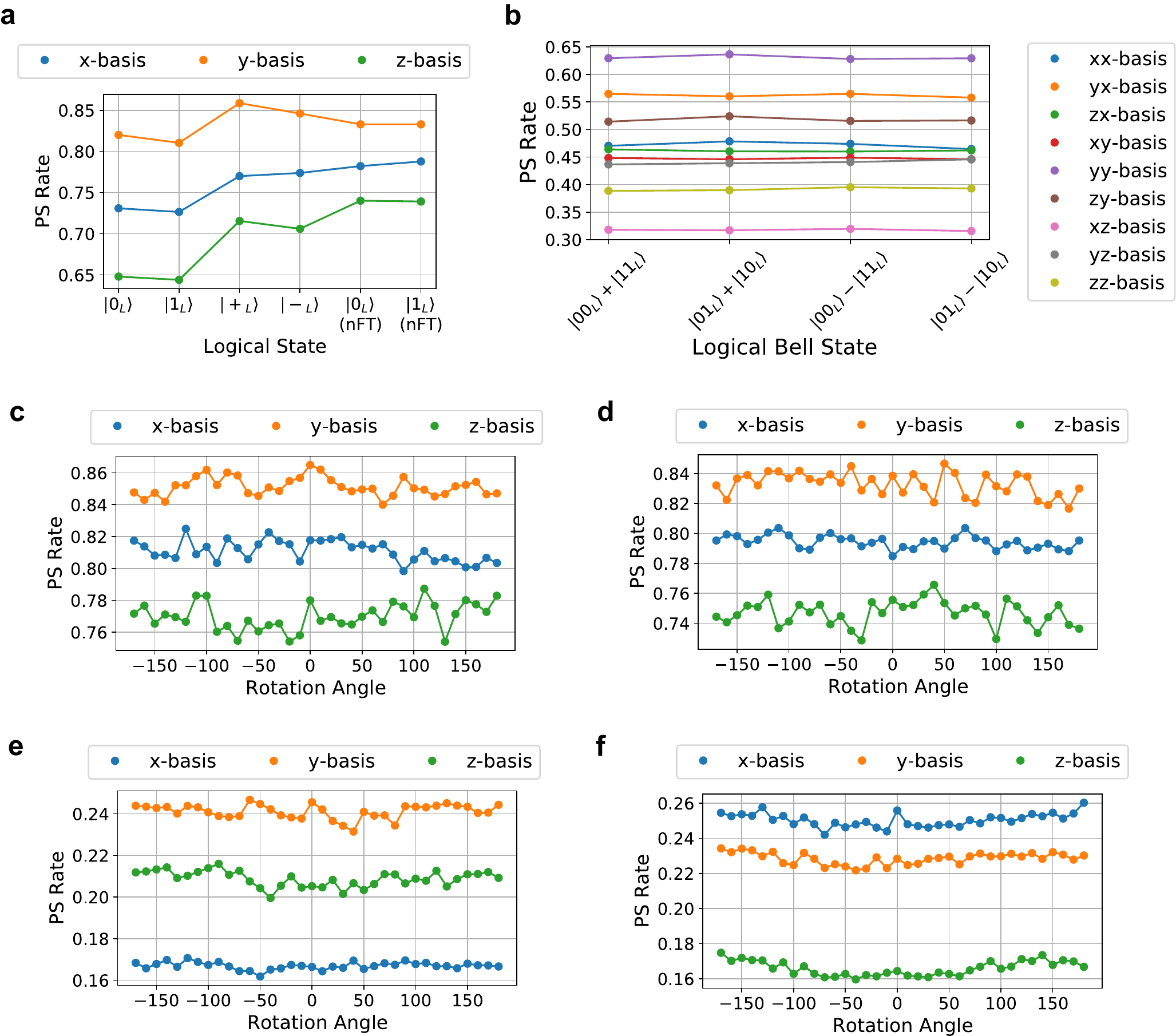}
\end{center}
\setlength{\abovecaptionskip}{0pt}
\caption{{\textbf{Data of post-selection rates under various measurement bases.} 
\textbf{(a)} Post-selection rates in single-qubit logical state preparations.
\textbf{(b)} Post-selection rates  four logical Bell state preparations.
\textbf{(c-d)} Post-selection rates of input states in single-qubit rotation experiments.
\textbf{(e-f)} Post-selection rates of output states in single-qubit rotation experiments.}}
\label{ps}
\end{figure*}}

\clearpage
\section{Numerical simulation}

Here, we numerically simulate the circuits in the main text by the Monte Carlo method. The depolarizing Pauli noise channels are used to simulate noise in the experiment circuits. Specifically, the depolarizing Pauli noise channels are defined as follows:

\begin{equation}
\begin{aligned}
\mathcal{E}_{1}(\rho_1)&=(1-p_1)\rho_1+(p_1/3)\sum_{P\in\{X,Y,Z\}}P\rho_1P,\\
\mathcal{E}_{2}(\rho_2)&=(1-p_2)\rho_2+(p_2/15)
\times \sum_{\substack{P_1,P_2\in\{I,X,Y,Z\},\\P_1\otimes P_2\neq I \otimes I}}P_1\otimes P_2\rho_2P_1\otimes P_2,
\end{aligned}
\end{equation}
where $\rho_1$ and $\rho_2$ are single-qubit and two-qubit density matrices respectively. Here we set $p_1$ for each qubit as the single-qubit gate error from Tab.~\ref{tables1}, and $p_2$ as the two-qubit gate infidelity from Tab.~\ref{tables2}. In the simulated circuits, we apply $\mathcal{E}_{1}$ after state initialization, single-qubit gates, and idle operations, and $\mathcal{E}_{2}$ after two-qubit gates. Additionally, The measurement result flips with a probability of $p_m=1- (F_{00} + F_{11}) / 2$, where $F_{00}$ and $F_{11}$ are from Tab.~\ref{tables1}. We simulate the circuits in the Heisenberg representation~\cite{gottesman1998heisenberg}, randomly inserting Pauli noise with corresponding probability. Since the circuits are Clifford circuits, according to the Gottesman-Knill theorem, they can be efficiently simulated in polynomial time~\cite{nielsen2010quantum,PhysRevA.70.052328}. 

\begin{figure*}[htbp]
\begin{center}
\includegraphics[width=0.7\linewidth]{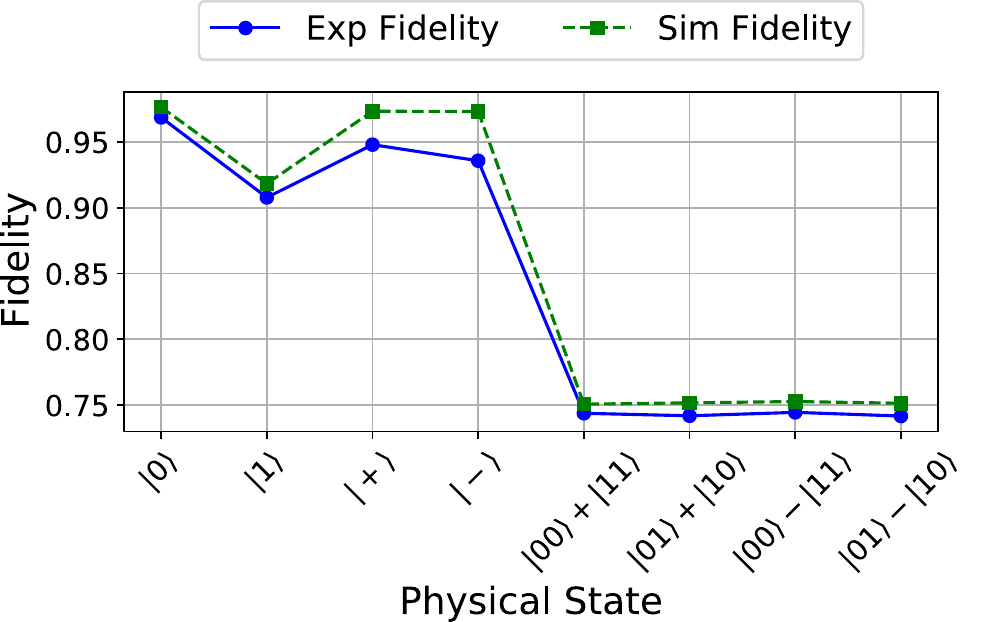}
\end{center}
\setlength{\abovecaptionskip}{0pt}
\caption{\textbf{Experimental and Simulated data of physical state fidelity.}}
\label{figs6}
\end{figure*}

Corresponding to the results in the main text, we simulated circuits of logical state preparation, logical Bell state preparation, and logical single-qubit rotations. Each circuit was sampled at least $10^6$ times. We performed the same post-selection process on the simulated and experimental data and calculate the fidelity. Similarly, we simulated the preparation and measurement circuits for single-qubit states and Bell states of the physical qubits, as shown in Fig.~\ref{figs6}. A slight difference is that the flip probability of measurement is \(p_m=1 - F_{00}\) if the measured state is the \(|0\rangle\) state, otherwise it is \(1 - F_{11}\). 

{To investigate the effects of different types of noise, circuits with only gate noise or measurement noise were simulated. Here, we focus on fault-tolerant preparation and measurement circuits of single-qubit states and Bell states, as only fault-tolerant operations can demonstrate the error detecting code's ability to suppress noise. When only gate noise is present, the two-qubit gate noise strength is set to $p_2=p$, and the single-qubit operation noise strength is $p_1=p/10$, with no measurement noise ($p_m=0$). When only measurement noise is present, $p_m=p$, and other noise types are set to zero. By observing the magnitude and slope of the error rates in Fig.~\ref{plus}~(a-b), we find that, in the experimental circuits we designed, measurement noise has a greater impact on the error rate, where the error rate is defined as the probability of obtaining an incorrect measurement result for the eigenvalue of a logical Pauli operator. Additionally, circuits with both gate and measurement noise were also simulated. We found that these circuits have a relatively high pseudo-threshold (over 10\%), significantly higher than the theoretical threshold of 1\% for surface codes. Here, the pseudo-threshold is defined as the value of $p$ at which the error rates of the logical and physical circuits are equal. This suggests that, for certain computational tasks, the logical qubits we designed are expected to outperform physical qubits under higher noise conditions, highlighting their potential application in early FTQC. These numerical simulation results are shown in Fig.~.\ref{plus}.}

{\begin{figure*}[t]
\begin{center}
\includegraphics[width=0.95\linewidth]{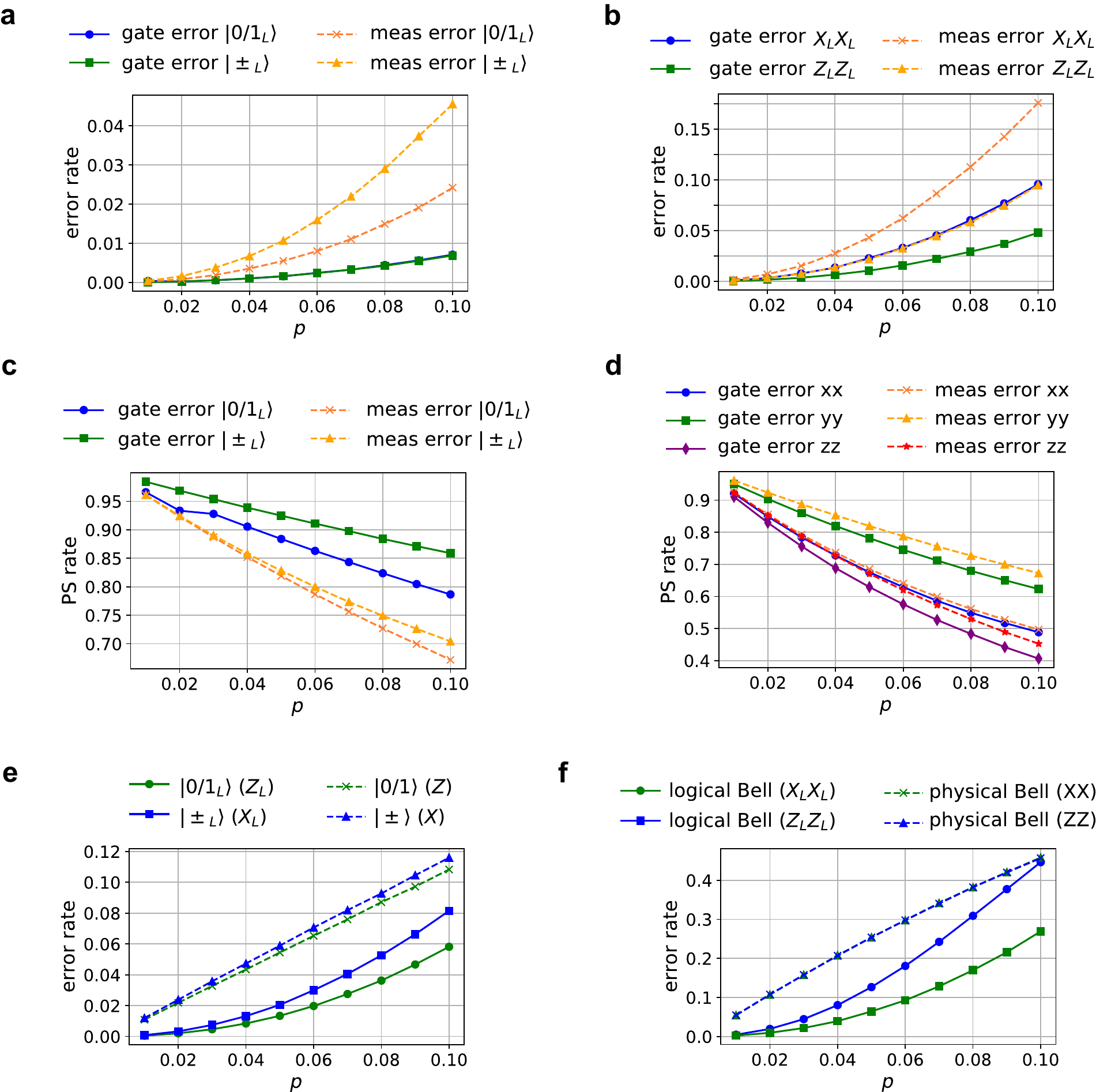}
\end{center}
\setlength{\abovecaptionskip}{0pt}
\caption{{\textbf{ Comparison of simulation results under the influence of different types of noise.} }
\textbf{(a-b)} Error rate of single-qubit logical states (a) or logical Bell states (b) with only gate error or measurement error as a function of $p$. We measured the eigenoperators of single-qubit logical states or the $X_LX_L$ and $Z_LZ_L$ operators of Bell states, which are distinguished in the legend.
\textbf{(c-d)} Post-selection rate under eigen-basis measurements of single-qubit logical states (c) or logical Bell states (d) with only gate error or measurement error as a function of $p$.  
\textbf{(e-f)} Error rate of single-qubit logical states (e) or logical Bell states (f) with both gate and measurement error present as a function of $p$.}
\label{plus}
\end{figure*}}

\begin{figure*}[t]
\begin{center}
\includegraphics[width=0.7\linewidth]{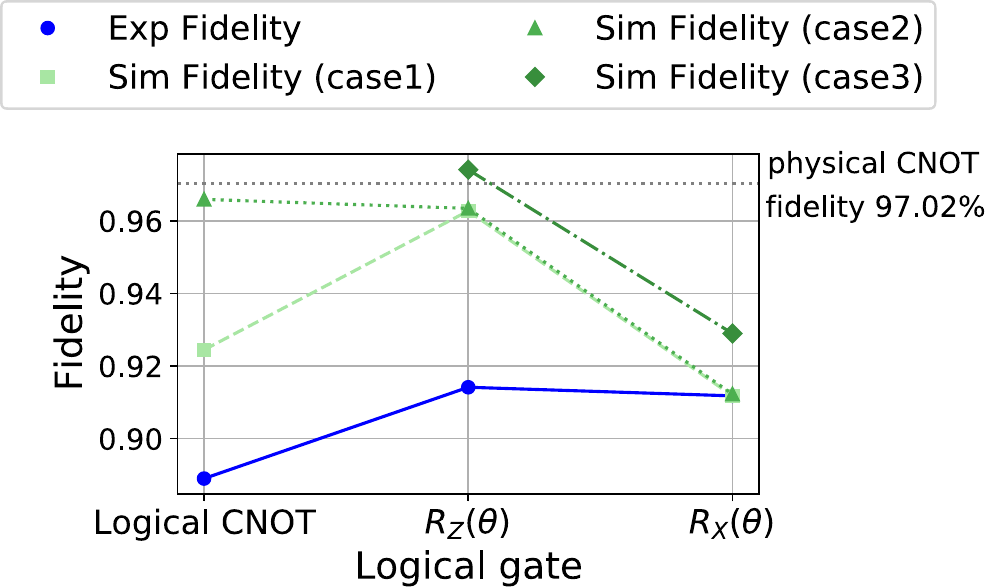}
\end{center}
\setlength{\abovecaptionskip}{0pt}
\caption{\textbf{Simulation results of the characterization of the logical CNOT gate and single-qubit rotation gates.} The fidelity of single-qubit rotation gates provided  are the average of $R_Z(\theta)$ or $R_X(\theta)$ with \( \theta\in\{0,\pi/4,\pi/2,\pi\}\).}
\label{figs7}
\end{figure*}

Lastly, we simulated the characterization circuits for the logical CNOT gate and logical single-qubit gates under three cases with different parameters. We processed the simulated data in the same way to construct LPTMs and thus obtained the fidelities of these logical gates. First, we simulated the circuit under the real experimental parameters (case 1). The dominant noise in our experiment comes from readout errors. As we used process tomography to characterize logical gates, the readout errors will significantly influence the characterization. To understand whether the current gate parameters have reached the pseudo-threshold point for logical gate implementation, we kept the gate parameters unchanged and set the measurements during the characterization part to be noiseless (case 2). The results showed that under these parameters, the fidelity of the logical CNOT gate approached that of the physical CNOT gate (97.02\%). We note that even after removing measurement noise, the LPTM still provides a conservative estimate of the logical gate fidelity because the logical $Y$ measurement is not fault-tolerant. Second, as the fidelity of ancilla states can be improved through distillation, we simulated the gate teleportation circuits again with noiseless ancilla state preparation (case 3), and found that average fidelity of all logical single-qubit gates still remained lower than that of the logical CNOT gate. This implies that at the logical qubit level, the fidelity of single-qubit gates is possibly lower than that of two-qubit gates implemented transversely, showing results contrary to those at the physical qubit level. These numerical simulation results are summarized in Fig.~\ref{figs7}.

{
\section{Uncertainty analysis}
In this paper, data uncertainty is determined using a binomial distribution to reflect the statistical fluctuations of the experimental results. For the binomially distributed data, uncertainty is calculated using the standard formula $\sigma = \sqrt{p(1 - p)/N}$, where $p$ is the probability of measuring a certain outcome and $N$ is the number of trials that passed the post-selection. To enhance the reliability of the results, a $2\sigma$ confidence interval is employed, ensuring approximately 95\% confidence in the reported values.}

{
\section{Extension of experiments and outlook}}

\begin{figure*}[htbp]
\begin{center}
\includegraphics[width=0.7\linewidth]{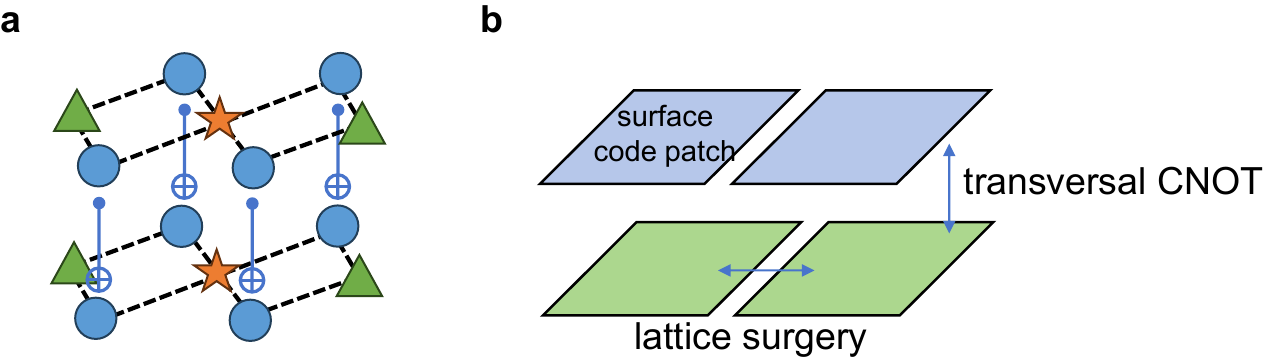}
\end{center}
\setlength{\abovecaptionskip}{0pt}
\caption{(a) General demonstration of the transversal CNOT gate in a two-layer architecture. Here, asterisks and triangles represent measurement qubits, while circles represent data qubits. The dashed lines between qubits indicate the topological connections within a single layer. Blue CNOT gates represent transversal CNOT operations between two layers.
(b) Lattice surgery and transverse CNOT gate hybrid scheme.}
\label{figs8}
\end{figure*}

{In the main text, logical qubits are encoded in a one-dimensional structure. However, when measurement qubits are added, this one-dimensional structure cannot be naturally maintained. For larger surface codes, although Ref.~\cite{PhysRevResearch.5.043137} achieved the initialization of distance-three surface codes on a one-dimensional chain, it remains unclear how to extend this to larger code distance. Based on these facts, in superconducting systems, a general demonstration of the transversal CNOT gate is more likely to be carried out on hardware with a two-layer structure. For example, Fig.~\ref{figs8}a illustrates the natural extension of our experiment.}

{In fact, superconducting hardware with a two-layer structure has not been the first to receive attention. For instance, Ref.~\cite{bravyi2024high} proposed a two-layer LDPC code, bringing superconducting platforms closer to realizing LDPC codes. This supports the demand for layered structures in superconducting hardware for quantum error correction and FTQC.}

{Implementing a hybrid scheme for FTQC on a two-layer superconducting platform is feasible. Logical operations between logical qubits in each layer can be achieved through lattice surgery, while transversal CNOT gates are allowed between corresponding logical qubits in different layers (see Fig.~\ref{figs8}b). This architecture effectively reduces the number of routing qubits required in lattice surgery. Typically, in lattice surgery, a single logical operation occupies a routing area of length $L$, where $L$ is the path length connecting two logical qubits. The path length $L$ determines the space-time overhead and error rate of the logical operation. Clearly, by using transversal CNOT gates between the two layers, the possible path length is effectively reduced, demonstrating the advantage of the hybrid approach that includes transversal gates.}

\end{document}